\documentclass[preprint,eqsecnum,aps]{revtex4}
\usepackage{graphicx}
\usepackage{dcolumn}
\usepackage{amsmath}

\makeatletter

\begin{document}

\title{The study of Anisotropic Flows at LHC with purturbative simulation}

\author{Ghi R. Shin}
\affiliation{Department of Physics, Andong National University,
Andong, South Korea}

\date{\today}
~ \\

\begin{abstract}
We study the harmonic flows, for example, the directed, elliptic, third and fourth flow of the system of partons formed
just after relativistic heavy ion collisions. We calculate the minijets produced during the primary collisions
using standard parton distributions for the incomming projectile and target nucleus. We solve the Boltzmann equations
of motion for the system of minijets by Monte Carlo method within only the perturbative sector.
Based on the flow data calculated, we conclude the simulation can not explain the experimental results
at RHIC and LHC so that the nonperturbative sector plays much more important roles even from the earliest
stage of heavy ion collisions.
\end{abstract}
\maketitle

\section{Introduction}
The Quark-Gluon Plasma (QGP)\cite{mul85}  is a fascinating state of matter, which is a QCD plasma consisting of quasi free quarks, 
antiquarks and gluons. This state of matter has been actively studied in theory and experiment. 
The Super Proton Synchrotron (SPS) at CERN had tried to produce the matter in the 1980s and 
Relativistic Heavy Ion Collider(RHIC) at BNL is continuing the search from 2000 and CERN's Large Hadron Collider(LHC) 
has joined the study from 2009. It is general consenseous that the QGP has been produced in laboratory.

One of reasons why so many efforts have been put on the study is the understanding of the theory of strong interaction,
Quantum Chromodynamics(QCD). The QCD is well known for that it is notoriously difficult to calculate any quantitative 
physical obserable from the theory: The static properties of the theory however can be obtained with lattice calculation and 
certain dynamical properties, for example, the hadron production from $e^- e^+$ scattering, can be estimated with
perturbative calculations but the detailed dynamical evolution of the QCD system in general can not be addressed with 
neither the lattice calculation nor the pertubative one up to date.
We thus have to keep in mind that the understanding of the QCD nonperturbative sector in dynamics is the primary goal
of the studying a QGP. 
The first question for that matter is 'how is the QGP formed?'. This highly non-trivial question should be one of major
problems we should answer. The next unanswered question is 'how is a hadron produced from the QGP dynamically?'.

We know that the 3+1D hydrodynamics\cite{schenke} incorporating viscous property is doing well 
to explain the evolution of a QGP. And the evolution of a hadronic gas is described well by hadron transport formalism, 
for example, uRQMD and so on. We however have to mention that the hydroformalism can not give any explanation
to those two critical questions. 

In order to study those unanswered problems it seems to us that it is best to use the quantum kinetic theory\cite{elze,geiger}.
Namely, we need to know how much the perturbative sector can explain the experiments
and how much the non-perturbative one should contribute.
This will help us to foumulate a realistic model for dynamical features.
Having this in mind we focus on the harmonic flows of the system
formed just after a heavy ion collision as a function of time with perturbative theory. 

The evolution of a heavy ion collision can be viewed in partonic point of view as follows;
A bunch of partons of a projectile nucleus(or nucleon) makes collisions with those partons of
target nucleus(or nucleon) to liberate the constituent partons to quasi free particles. Those freed partons can radiate
photons and partons and make collisions with other partons. During this period of evolution, the system may
reach thermal and chemical equilibrium to form the QGP. The QGP will then produce(or convert into) hadrons 
eventually after expanding sufficient enough to break up. This hadronic gas will further evolve.
We follow this view point as much as we can in our study.

We calculate the primary partons which are liberated by collisions between projectile and target nucleus in Section III
and briefly describe our numeric procedure which is a parton evolution code in Section IV.
We present the simulation results and discussion in Section V and conclude in Section VI.

\section{Primary minijet production}
Assuming that partons are produced in relativistic heavy-ion collisions
by elastic scattering between the constituents of a projectile nucleus and those of a target nucleus,
we can write the total number of collision events \cite{eichten, ham99, nayak, eks02,coo02}:
\begin{eqnarray}
{N^{event}} &=& K T(b) \int dy_3 dy_4 d^2 p_T 
\sum_{ij,\,kl} [ x_1 f_{i/A}(x_1,Q_0^2) x_2 f_{j/B}(x_2,Q_0^2)
{{d \sigma^{ij\rightarrow kl}(\hat{s},\hat{t},\hat{u})} \over {d {\hat t}}} \nonumber  \\
& & + x_1 f_{j/A}(x_1,Q_0^2) x_2 f_{i/B}(x_2,Q_0^2) {{d \sigma^{ij\rightarrow kl}(\hat{s},\hat{u},\hat{t})}
\over {d {\hat t}}}  ] { 1 \over {1+ \delta_{ij} }},
\label{pt_y_dis}
\end{eqnarray}
where we assume that there are no correlations between the momentum 
and space coordinates of a constituent parton.
We explicitly neglect the transversal momentum of incomming partons and 
$s$ is the CM energy squared of two mother nucleons and $b$ is an impact parameter.
$K$ is the K-factor to include the higher-order diagrams; we will set $K = 2$ for the RHIC energy and $1.5$ for LHC energy.
A parton $i$ of nucleus $A$ collides with a parton $j$ of nucleus $B$ and produces
partons $k$ and $l$ or a parton $j$ of nucleus $A$ collides with a parton $i$ of nucleus $B$ to produce
partons $k$ and $l$. 
Each parton has rapidity $y_1$, $y_2$, $y_3$, and $y_4$, respectively. $x_1$ and $x_2$ are
the Bjorken scaling variables of incomming partons.
The relations between the variables before and after the collision are given by
\begin{eqnarray}
x_1 &=& p_T (e^{y_1} + e^{y_2}) /\sqrt{s},\\
x_2 &=& p_T (e^{-y_1} + e^{-y_2}) /\sqrt{s},\\
\hat{s} &=& x_1 x_2 s, \\
\hat{t} &=& -p_T^2 (1 +e^{y_2 - y_1}),\\
\hat{u} &=& -p_T^2 (1 +e^{y_1 - y_2}).
\end{eqnarray}
We write the available kinematic region for convenience \cite{ham99, shin02, shin03},
\begin{eqnarray}
Q_0\, ^2 \leq p_T\, ^2 \leq ({{\sqrt{s}}\over{2 \cosh y}})^2, \\
-\log ({{\sqrt{s}}\over{p_T}} - e^{-y}) \leq y_4 \leq 
\log({{\sqrt{s}}\over{p_T}} - e^{-y}),\\
|y| \leq \log ( {{\sqrt{s}}\over{2 Q_0}} + \sqrt{{{s}\over{4Q_0\,^2}} - 1}).
\end{eqnarray}

$Q_0$ is the momentum scale which we are probing the nucleus and is a minimum momentum transfer.
The hat on the Mandelstam variables means that those are the variables of a parton.
The processes we consider in our study are $gg \leftrightarrow gg + q \bar q$, 
$g q \leftrightarrow gq$, $g \bar q \leftrightarrow g \bar q$, $q^a q^b \leftrightarrow q^c q^d $, 
$q\bar q \leftrightarrow q \bar q$, $\bar q^a \bar q^b \leftrightarrow \bar q^c \bar q^d $.
We do not include some basic channels, such as $qg \rightarrow q \gamma$, 
$q \bar q \rightarrow \gamma \gamma$, and $q \bar q \rightarrow g \gamma$,
which could provide important information on the system. 

Assuming the nucleons in a nucleus are treated independently except the shadow effect, 
we can write the parton distribution of the nucleus A as follows:
\begin{eqnarray}
f_{i/A}(x,Q^2) = f_{i/N}(x,Q^2) R_A(x,Q^2),
\end{eqnarray}
where $f_{i/N}(x,Q^2)$ is the parton distribution of a free nucleon and 
$R_A(x,Q^2)$ is the nucleus ratio function, which is the nucleon distribution of the nucleus, $A$.
We use the CTEQ4\cite{cteq} or the GRV98\cite{grv98} distribution function for a free nucleon parton distribution 
and the EKS98 parametrization for the ratio function\cite{eks99}. 

Assuming that the density of the nucleus
is constant over the sphere of radius $R$ with the sharp edge, we can define the nuclear thickness function,
\begin{eqnarray}
\tau(\vec r) &=& \int_V dz \rho(\vec r, z),
\end{eqnarray}
where $\vec r$ is the transversal vector and 
\begin{eqnarray}
\tau_A(\vec r_A) &=& 2 \rho_0^A \sqrt{R_A^2-(\vec r - \vec b/2)^2},\\
\tau_B(\vec r_B) &=& 2 \rho_0^B \sqrt{R_B^2-(\vec r + \vec b/2)^2},
\end{eqnarray}
where $\vec b$ is the impact parameter, the vector from the center of a target B to
the center of a projectile A.

The overlap function between the target and projectile nucleus is then
\begin{eqnarray}
T_{AB} (\vec r ;\vec b ) &=& \tau_A(\vec r - \vec b /2 ) \tau_B(\vec r + \vec b/2),\nonumber \\
&=& 4 \rho_0^A \rho_0^B \sqrt{R_A^2-(\vec r - \vec b/2)^2}
\sqrt{R_B^2-(\vec r + \vec b/2)^2},
\label{overlap_fn}
\end{eqnarray}
and the nuclear geometric factor  $T(b)$ at a given impact parameter $b$ is
\begin{eqnarray}
T(b) &=& \int d{\vec r} \,\, T_{AB}(\vec r, \vec b ).
\end{eqnarray}

We can choose the (transversal) position according to the overlap function, Eq.\ref{overlap_fn}.
To sample the position according to this probability density can be performed
using Veto algorithm, namely we sample $(x,y)$ within the
allowed region randomly and calculate the probability density at that position
to give $P(x,y;\vec b)=T_{AB} (\vec r ;\vec b ) $. We generate a random number $r$ and compare
to $P(x,y;\vec b)/P(0,0;\vec b)$. If $r$ is less than $P(x,y;\vec b)/P(0,0;\vec b)$,
we accept the $(x,y)$ but if $r$ greater than that, we reject and sample another
positions.  

To obtain the longitudinal position and the collision time of a collision, we consider two nonrelativistic classical balls
passing through each other and choose $t=0$ as the time when
two colliding nucleus make first contact at the impact parameter $b=0$ in CM frame.
At the given transverse collision position $(x,y)$, we consider a longitudinal tube
of transversal area $\delta A$ through the point $(x,y)$ with the half thickness $D_A=\sqrt{R_A^2 -r_A^2}$
of nucleus $A$ and $D_B=\sqrt{R_B^2 -r_B^2}$ of nucleus $B$ where
$r_A^2 = (\vec r - \vec b /2)^2$ and $r_B^2 = (\vec r + \vec b/2)^2$.
The tubes from both spheres, of which have different lenght from each other,
begin to overlap one another starting at 
$t_s = {{(R_A-D_A)+(R_B-D_B)}\over {v_A+v_B}}$  and completely overlap at
$t_1 = t_s + l_{min}/(v_A+v_B)$ and begins to seperate at
$t_2 = t_s + l_{max}/(v_A+v_B)$ and completely seperate at
$t_e = {{(D_A+D_B)}\over{v_A+v_B}} + t_s$, where $v$ is the velocity of a sphere and
$l_{min}$ and $l_{max}$ are the minimum and maximum of $D_A$ and $D_B$.

Assuming the probability density which an elastic collision can be occurred
is proportional to the overlap volume, we have the probability density of collision,
\begin{eqnarray}
P(t) &=& (t-t_s) (v_A+v_B) ,	\,\, t_s < t < t_1,\\
&=& l_{min},	\,\,\,\, t_1 < t < t_2, \\
&=& (t_2 - t) (v_A+v_B),	\,\, t_2 < t < t_e ,
\end{eqnarray}
so that the overall probabilities are
\begin{eqnarray}
S_1 &=& {1 \over 2} (t_1-t_s) l_{min} , \\
S_2 &=& (t_2 - t_1 ) l_{min},	\\
S_3 &=& {1 \over 2} (t_e - t_2) l_{min} .
\end{eqnarray}
We can generate a random number $r_1$ and choose the region depending on the overall
probabilities. Once we have a region, we can generate the second random number, $r_2$,
and the collision time is given by,
\begin{eqnarray}
t&=& t_s + \sqrt{ S_1 r_2 v} 
\end{eqnarray}
for the region $S_1$, and
\begin{eqnarray}
t&=& t_1 + (t_2 - t_1) r_2, \\
t &=& t_e - \sqrt{t_e^2+t_2^2 - 2 t_e t_2 - S_3 v r_2 }, 
\end{eqnarray}
for resion $S_2$ and $S_3$ respectively.

Once we have the collision time $t$, we can choose the longitudinal collision position $z$ within the overlap tube
by using a Monte Carlo method:
\begin{eqnarray}
z &=& z_c - (t-t_s)v + 2 v t r,
\end{eqnarray}
if $t_s < t < t_1 $,
\begin{eqnarray}
z &=& z_c - (t-t_s)v + l_{min} r, \\
z &=& z_c - D_A + 9t-t_s)v + [l_A+l_B-2(t-t_s)v] r,
\end{eqnarray}
for $t_s < t < t_1 $ and $t_2 < t < t_e $ respectively, where $r$ is a random number.
$z_c = {{(R_B-D_B)-(R_A-D_A)}\over 2}$ is the center of the first contact points at $t=0$.

We can apply the argument to a relativistic collision. Consider a sphere of $v \sim c$. 
The sphere is then contracted by the $\gamma$-factor so that
the starting and the ending collision times for the identical spheres in the CM frame
are $t_s = (R-D)/\gamma$ and $t_e = (R+D)/\gamma$,
respectively, where $D=\sqrt{R^2-x^2-y^2}$. 
The collision time probability is, thus, given by
\begin{eqnarray}
P(t) &=& {{\gamma^2}\over{2D^2}}(t-t_s)
\end{eqnarray}
so that the collision time and the longitudinal position can be obtained from a Monte Carlo method:
\begin{eqnarray}
t &=& t_s + {{2D}\over{\gamma}}\sqrt{r_1}, \label{rand_t} \\
z &=& 2(t-t_s) r_2 - (t-t_s),
\end{eqnarray}
where $(r_1,r_2)$ are random numbers between 0 and 1.\\

The momentum of a produced parton can be obtained from the minijet distribution, Eq. (\ref{pt_y_dis}):
\begin{eqnarray}
f(p_T, y) &=& C {1 \over p_T^2} {{dN^{jet}}\over{dy \,dp_T}},
\end{eqnarray}
where $C$ is a normalization constant.
We can choose $(p_T,y)$ for each test particle with this distribution function
by using a Monte Carlo sampling method. 
On the other hand, the azimuthal angle $\phi$ of the momentum
can be chosen with equal weight between $(0,2\pi)$.
These give the energy-momentum of the produced parton which is on-mass shell.\\

\begin{figure}[htp]
\centering
\includegraphics[width=100mm]{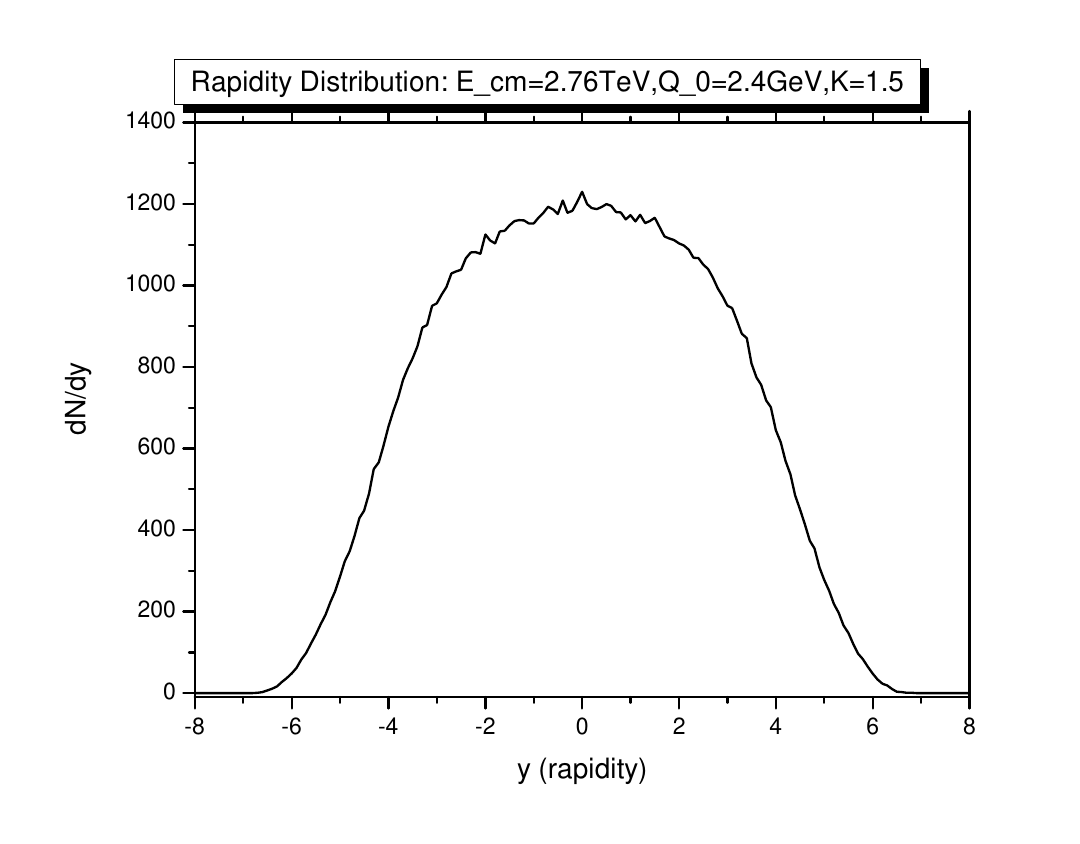}
\caption{Rapidity distribution of the test partons sampled
by a Monte Carlo method}
\label{fig1}
\end{figure}

\begin{figure}[htbp]
\centering
  \begin{minipage}[t]{3 in}
   \includegraphics[width=1.0\textwidth]{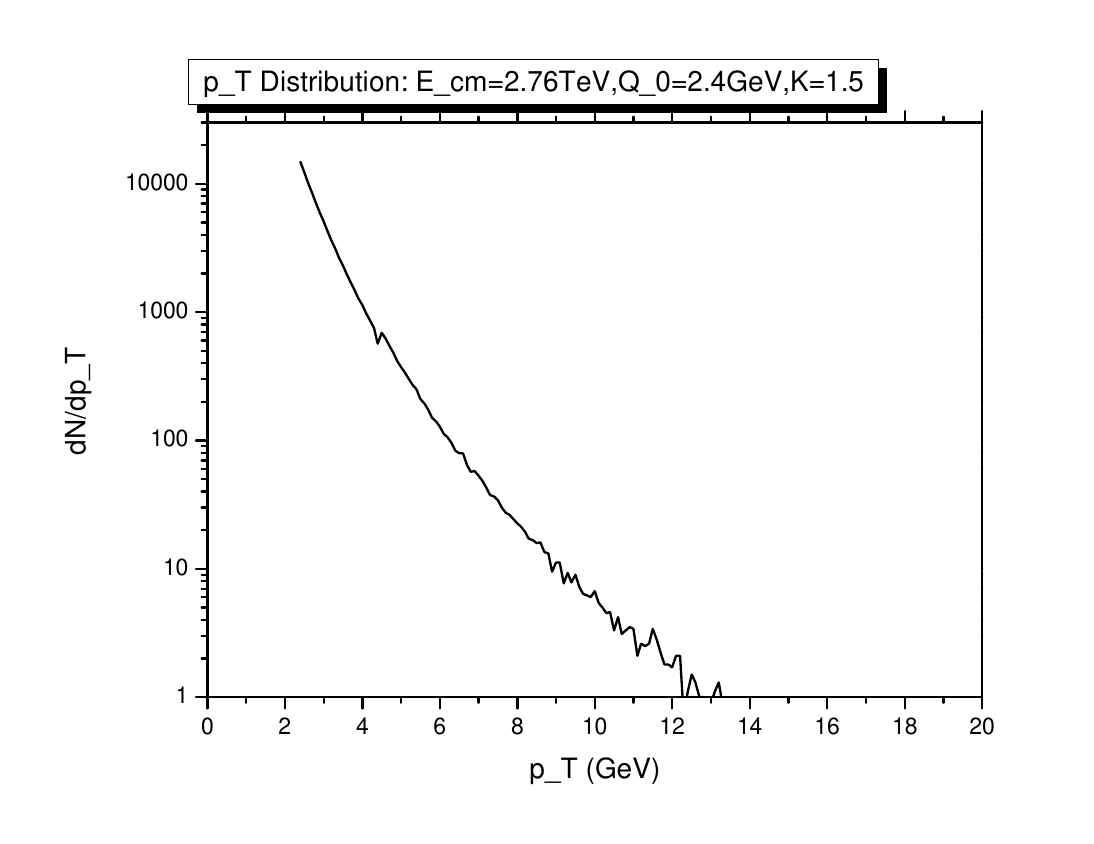}
    \caption{$p_T$ distribution of the test partons sampled
by a Monte Carlo method}
    \label{fig2}
\end{minipage}
\hspace{.1in}
\begin{minipage}[t]{3 in}
  \includegraphics[width=1.0\textwidth]{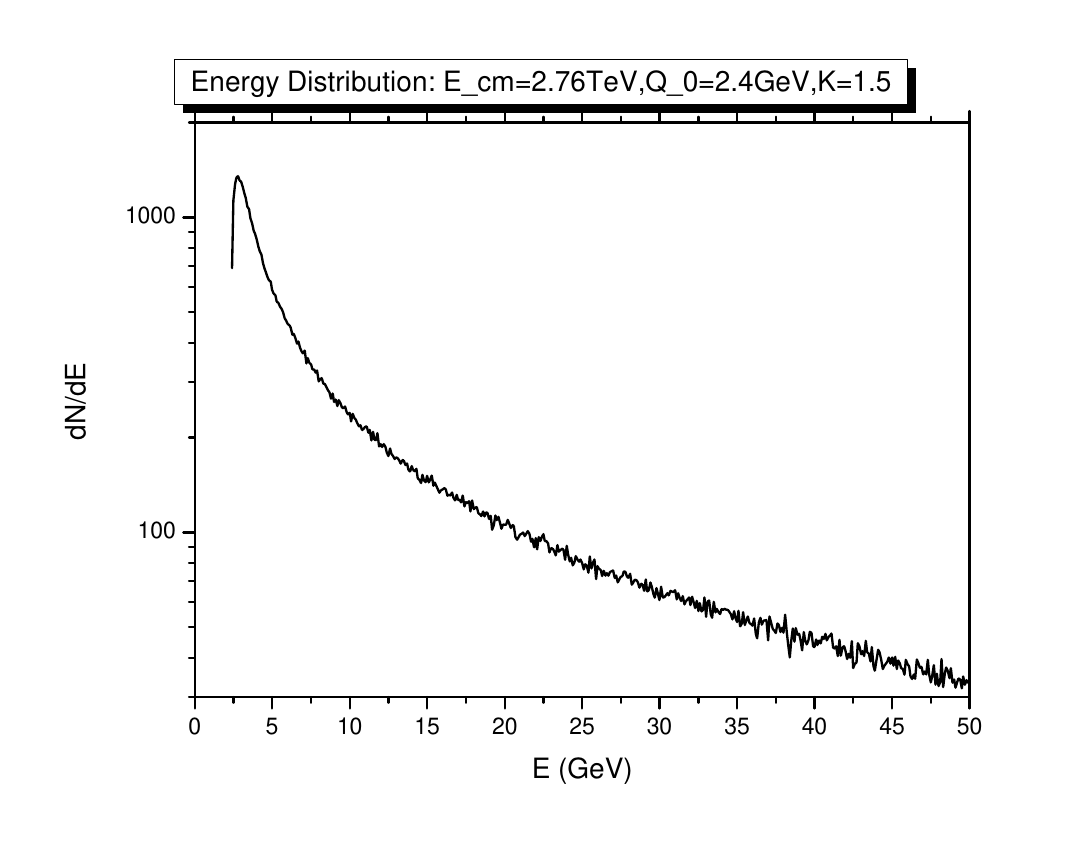}
  \caption{Energy distribution of the test partons sampled
by a Monte Carlo method}
  \label{fig3}
\end{minipage}
\end{figure}

The number of minijets depends on the probing momentum $Q_0$ and K-factor. 
We assume that the total energy of produced minijets is about 70 - 75$\%$ of the total CM energy.
This gives us $Q_0=2.4 \, GeV$ with $K=1.5$ for CTEQ4 at LHC energy $\sqrt s = 2.76 \, TeV$ and
the number of minijets is about 9,700 partons for $b=0$, which are mostly gluons. 
At RHIC energy, $Q_0=1.6 \, GeV$ and $K=2$ to produce 3800 partons.
Fig. \ref{fig1} shows the rapidity distribution of the Monte-Carlo-sampled test particles.
The distribution shows that the rapidity is almost flat at the central region, but falls quickly off with $|y| > 2.5$.
Figs. \ref{fig2}, \ref{fig3} show $p_T$ and energy distribution at LHC energy.

\section{Parton Evolution Simulation}

The evolution of a system of partons can be best described by the quantum transport equations\cite{elze, geiger}
based on the field theory. But it is too complicated to solve even numerically.
The semi-classical Boltzmann equations of motion thus are used to capture the main physics\cite{gei92}.
To solve the partonic Boltzmann equations, we use partonic Monte Carlo simulation(PCC)\cite{shin02,shin03}
which implements the main features of perturbative QCD, 
which includes the gluon radiation ($gg \rightarrow ggg$) channel in the secondary collision.
The algorithm of the simulation is simple and straightforward: A parton, which is a minijet and was produced
from the primary collisions between projectile and target nucleus(or nucleon), is following the straight classical 
trajectory depending on the initial momentum and position until it hits other parton. 
In order to decide whether the two particles makes a collision or not, we calculate the impact parameter
or the closest distance, $r_{min}$, between two particles and compare the distance with the radius of cross section,
$r_c=\sqrt{\sigma/\pi}$.
If $r_{min} < r_c$, those two have scattering. This decision making is rather deterministic
even though it should be probabilistic in quantum nature. We however note that although the decision is deterministic,
the outcomes, for example, scattering channels and energy-momentum of outgoing particles, are
stochastic and Markobian. Namely we choose the scattering channel out of many possible ones
based on the probabilistic weight with no history. Once the chennel is chosen, we sample by Monte Carlo
the outgoing momentum according to the differential cross section. In this way the parton system evolves
up to the time set by outside.

In this study, the small angle scatterings between test partons,
$sin \theta \geq p_{min} / E$ where $p_{min}$ is the minimum momentum transfer, is set at $p_{min} = 0.3 GeV/c$.
We put the QCD coupling constant to be $\alpha_s = 0.3$ throughout the simulation. 
The realistic value of K-factor is 1 to 2 to include the higher-order diagrams, but we will set
$K=2$ or $K=20$.
We know these values of basic parameter for the simulation are beyond the limit of perturbative calculation
in some case. The idea of this setting is to get the maximal outcomes from the perturbative sector.

The processes we consider in this study are $gg \leftrightarrow gg + q \bar q$, 
$g q \leftrightarrow gq$, $g \bar q \leftrightarrow g \bar q$,
$q^a q^b \leftrightarrow q^c q^d$, $q\bar q \leftrightarrow q \bar q$,
$\bar q^a \bar q^b \leftrightarrow \bar q^c \bar q^d$, and $gg \rightarrow ggg$ .
The cross sections for the processes up to the leading order (LO) can be found in Ref. \cite{peskin};
the total cross section of $gg \rightarrow gg$ with the momentum cutoff, for example, is about $10/GeV^2$,
which is about 4 $mb$. 

\section{Simulation results and discussion}

The primary partons, which are produced directly from the colliding nuclei, have azimuthal symmetry
in the momentum space since the collision cross section between partons is independent of azimuthal angle.
The partons will have numerous collisions among themselves after they are born and
may develop anisotropy in the momentum distribution if there is spatial anisotropy.
This azimuthal anisotropy, which is called the harmonic flow collectively, provides very important information
of the system. 
This azimuthal anisotropy can be extracted systematically by expanding the number density
as a function of azimuthal angle, $\phi$, with respect to the reaction plane\cite{oll},
\begin{eqnarray}
E {{d^3 N} \over {dp^3}} &=& {1 \over \pi} {{d^2 N} \over {dp_T^2 dy}} [ v_0 + 2 v_1 cos \phi
+ 2 v_2 cos 2 \phi + ... ],
\end{eqnarray}
where $v_1$ is the directed flow and $v_2$ the elliptic flow. When a parton's momentum is known, 
the directed and elliptic flow can be calculated by using the equation
\begin{eqnarray}
v_1 &=& < cos \phi > \, =\, <{{ p_x} \over {\sqrt{p_x^2 + p_y^2 }}} > \\
v_2 &=& < cos 2 \phi > \, =\, < {{p_x ^2 - p_y ^2 } \over { p_x ^2 + p_y ^2 }} >,
\end{eqnarray}
where the bracket denotes the average over the partons, and the impact parameter vector and collision
axis define the reaction plane. 
The directed flow tell us the sideward motion of particles in heavy ion collisions and it carries information
developed the earliest stage of collisions. It is argued that the directed flow could reveal a signature of a 
possible phase transition from normal nuclear matter to a QGP\cite{direct}.
The elliptic flow is a fundamental observable and is known as one of probes of QGP formation.
It reflects how the initial spatial anisotropy of the nuclear overlap region of primary nuclei collision is translated 
into the asymmetric momentum distribution of final particles.

\begin{figure}[htp]
\includegraphics[width=100mm]{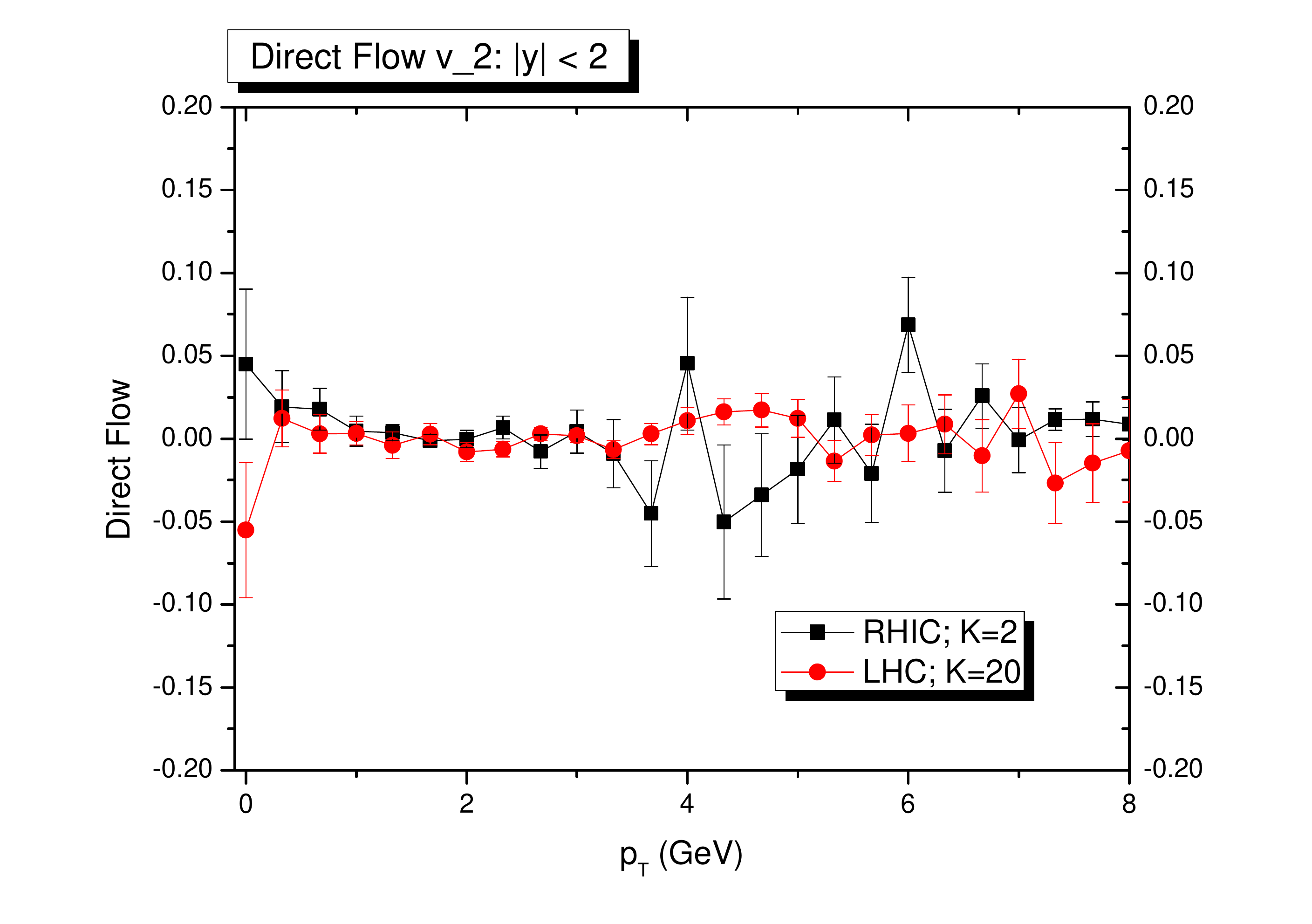}
\caption{Directed flow for RHIC and LHC at $b=7$ fm.} 
\label{fig4}
\end{figure}
Figure \ref{fig4} shows the directed flow as a function of transverse momentum
for the parton system. The data have been obtained by averaging over 100 runs.
We include only partons with $ |y| < 2 $. The impact parameter is $b=7 fm$ which corresponds to
20 - 30 \% in centrality\cite{centrality}. There are big error bars for small and large $p_T$
since the number of partons in the region is small. The simulation shows no directed flow
while significant amount of $v_1$ have been reported in experiments \cite{direct2}.
Here the RHIC energy means $ \sqrt{s} = 200 \,\rm GeV$ per pair of nucleons and 
LHC energy $ \sqrt{s} = 2.76 \,\rm TeV$ per pair.

\begin{figure}[htp]
\includegraphics[width=100mm]{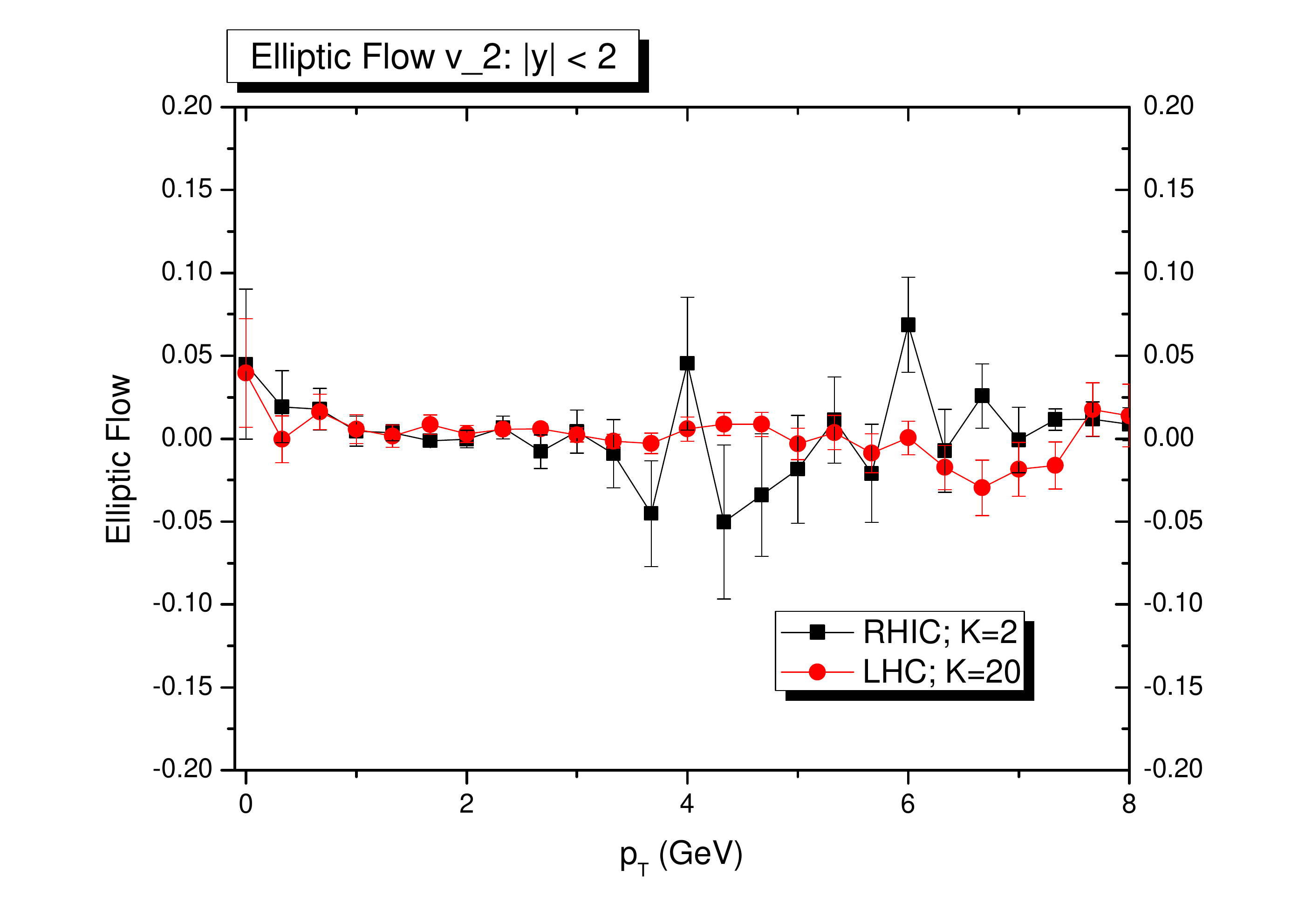}
\caption{Elliptic flow for RHIC and LHC energy at $b=7$ fm.} 
\label{fig5}
\end{figure}
Figure \ref{fig5} shows the elliptic flow as a function of transverse momentum
for the parton system. The simulation data show both energy have no elliptic flows
while ALICE\cite{alice} reported the elliptic flow as big as RHIC\cite{brahms05,phobos05,star05,phen05}. 
This is particularly interesting; 
even though we use much bigger perturbative cross sections than the reasonable ones
by setting $K=20$, we have null elliptic flows over the transverse momentum.
This clearly shows the failure of naive perturbative calculation.

\begin{figure}[htp]
\includegraphics[width=100mm]{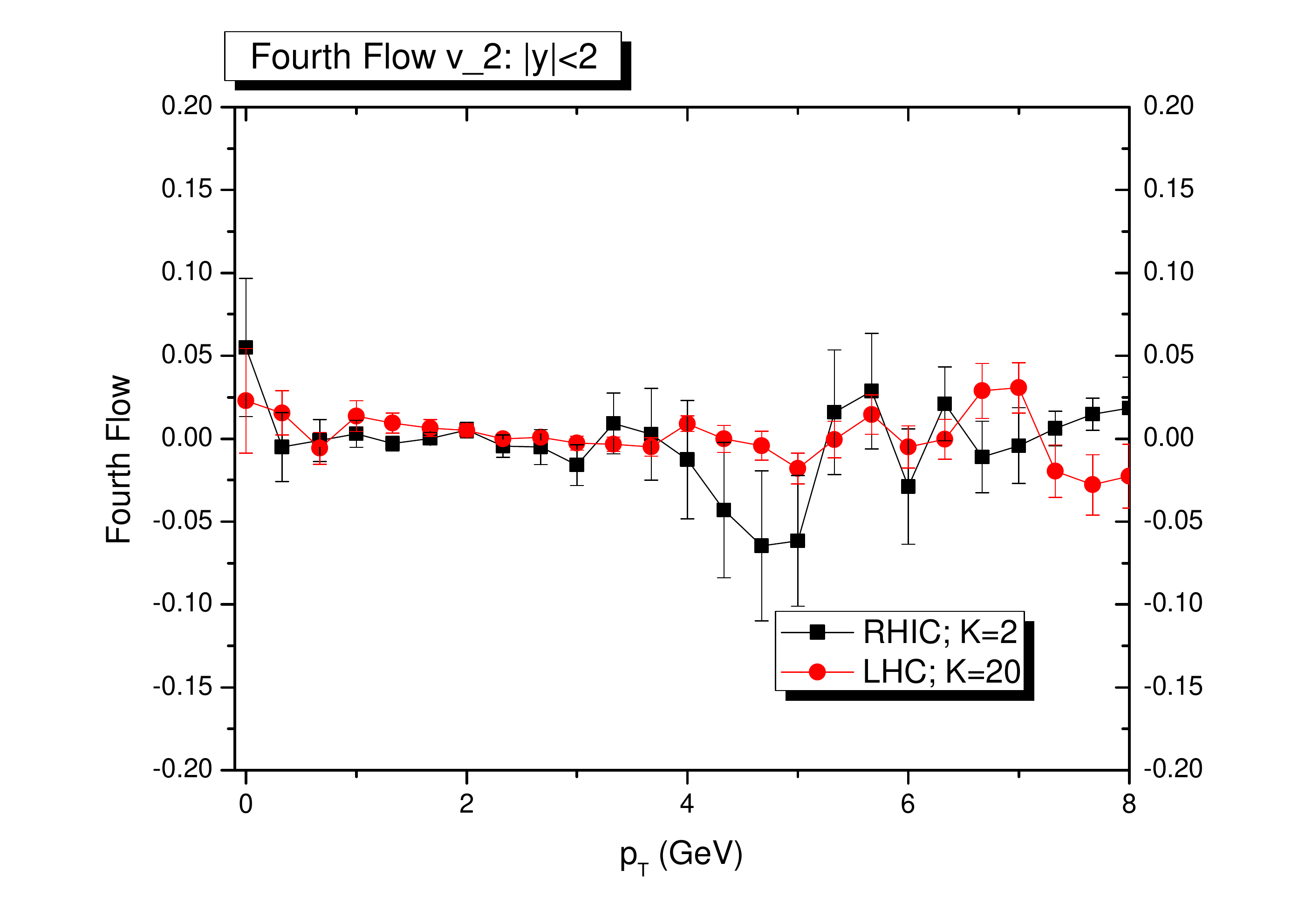}
\caption{Fourth flow for RHIC and LHC at $b=7$ fm.} 
\label{fig6}
\end{figure}
Figure \ref{fig6} shows the fourth harmonic flow as a function of transverse momentum
for the parton system formed just after heavy ion collisions. 

\begin{figure}[htp]
\includegraphics[width=100mm]{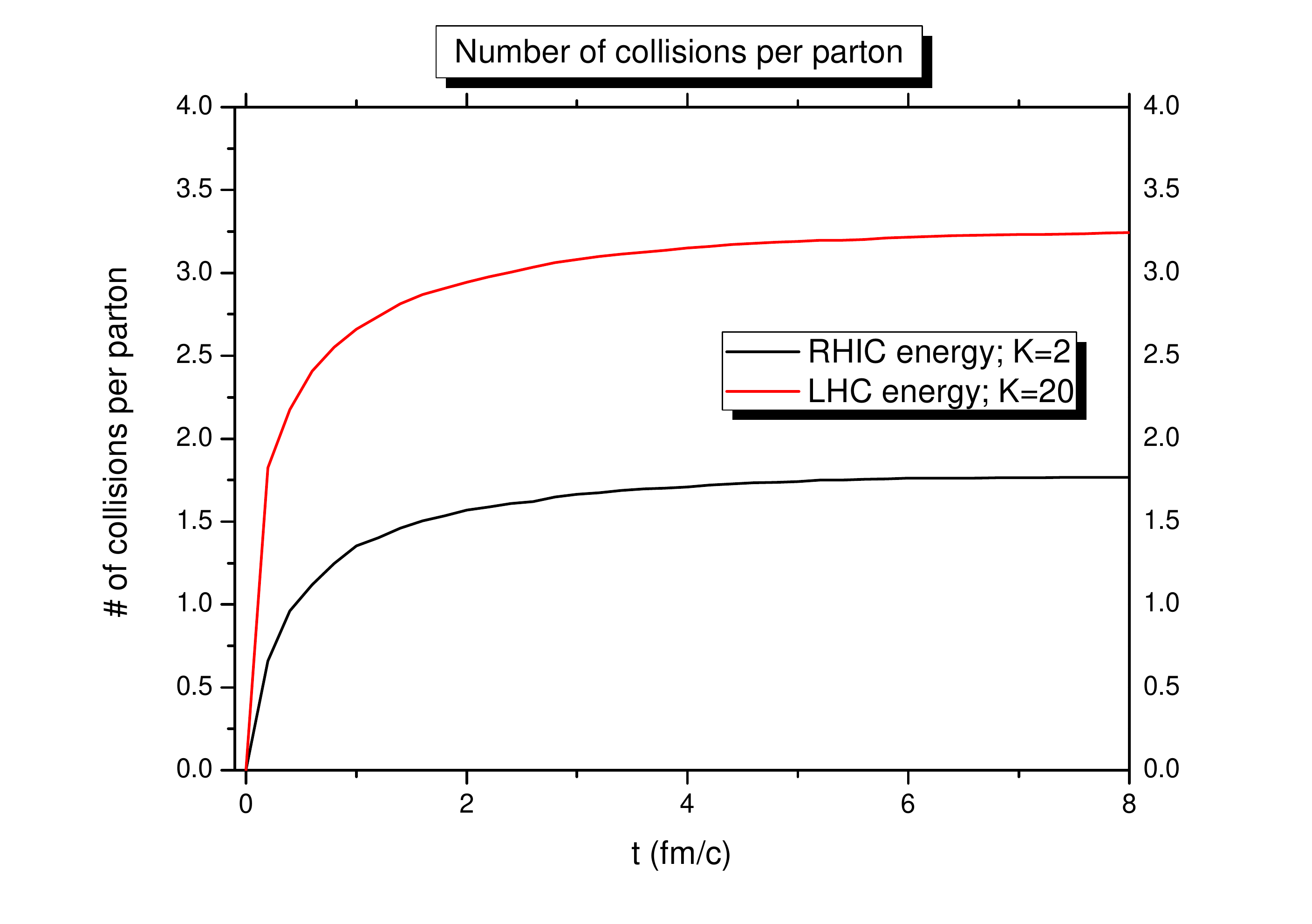}
\caption{Number of collisions per parton at RHIC and LHC energy at $b=7$ fm.} 
\label{fig7}
\end{figure}

To understand the flow data of simulations further, we calculate the number of collisions
as a function of time. Figure \ref{fig7} shows the cumulative number of collisions per parton.
The figure tell us that  the increasing rate at early stage is much higher at LHC than at RHIC
and perturbative collisions occur within 2 fm/c in both cases. This is interesting because
this tells us that all happen within 1-2 $fm/c$ and then the system is free streaming no matter
what the density is. 
We also calculated the number of collision at LHC with $K=2$ and found that 
the number is almost same as that of $K=20$. We can understand this as follows; 
The available phase space, especially the momentum space, at LHC is so large that 
the number of collision per parton do not increase as expected as the cross section does.

\section{Conclusions}
We study relativistic heavy ion collisions with Monte Carlo simulation: Using CTEQ4 and GRV98
we obtain the parton phase space distribution of nuclei (Au and Pb). When two colliding nuclei
overlap, the constituent partons make collisions and are freed if the momentum transfer is
greater than $Q_0$. Those partons formed just after primary collision evolve further.
We stop the evolution at $t=10 fm/c$ and analyze the results, especially the harmonic flows.
All the simulation data is compelling us that the naive perturbative calculation cannot explain the results
of RHIC and LHC. We further calculated the simulations using GRV98 distribution for a proton and  
found no difference from what we concluded here.

There are two places to improve the perturbative sector, which are not implemented yet in our study:
The first one is parton radiation. 
The partons of a system are off-mass shell and are surely subject of parton shower, namely
$1 \rightarrow 2$ branching. These will introduce many new partons into the system and 
could increase the number of collisions and flow effects.
The other important component which is missing
in our simulations is a color electromagnetic force\cite{mro}. 
We expect this color force will not change the harmonic flows since the force does not depend
on the azimuthal angle of partons but could excel the parton's thermal equilibration.
These two issues are under investigation by us.\\

{\it{$^*$Acknowledgements}}: This research was supported by Basic Science Research Program
through the National Research Foundation of Korea(NRF) funded by the Ministry of Education, 
Science and Technology(2010-0022228)

\end{document}